\documentclass[aps,prd,twocolumn,floatfix,superscriptaddress]{revtex4-2}

\usepackage{tabularx}
\usepackage{graphicx}
\usepackage{amsmath}
\usepackage{amssymb}
\usepackage{xspace}
\usepackage{multirow}
\usepackage{natbib}
\usepackage{hyperref}

\newcommand{\firstpaper}{P1\xspace}
\newcommand{\fitalgoname}{GBSIEVER\xspace}
\newcommand{\Taijimath}{\rm Taiji-mod\xspace}
\newcommand{\gbamp}{\mathcal{A}}
\newcommand{\intpar}{\kappa}
\newcommand{\fstat}{$\mathcal{F}$-statistic\xspace}

\DeclareMathOperator*{\argmin}{argmin}
\DeclareMathOperator*{\argmax}{argmax}

\begin{document}
\title{Resolving Galactic binaries using a network of space-borne gravitational wave detectors}
\author{Xue-Hao Zhang}
\affiliation{Lanzhou Center for Theoretical Physics, Key Laboratory of Theoretical Physics of Gansu Province, School of Physical Science and Technology, Lanzhou University, Lanzhou 730000, China}
\affiliation{Institute of Theoretical Physics \& Research Center of Gravitation, Lanzhou University, Lanzhou 730000, China}
\affiliation{Morningside Center of Mathematics, Academy of Mathematics and System Science, Chinese Academy of Sciences, 55, Zhong Guan Cun Donglu, Beijing 100190, China}
\author{Shao-Dong Zhao}
\affiliation{Lanzhou Center for Theoretical Physics, Key Laboratory of Theoretical Physics of Gansu Province, School of Physical Science and Technology, Lanzhou University, Lanzhou 730000, China}
\affiliation{Institute of Theoretical Physics \& Research Center of Gravitation, Lanzhou University, Lanzhou 730000, China}
\affiliation{Morningside Center of Mathematics, Academy of Mathematics and System Science, Chinese Academy of Sciences, 55, Zhong Guan Cun Donglu, Beijing 100190, China}
\author{Soumya D.~Mohanty}
\affiliation{Dept.~of Physics and Astronomy, University of Texas Rio Grande Valley, One West University Blvd.,
Brownsville, Texas 78520, USA}
\affiliation{Morningside Center of Mathematics, Academy of Mathematics and System Science, Chinese Academy of Sciences, 55, Zhong Guan Cun Donglu, Beijing 100190, China}
\email{soumya.mohanty@utrgv.edu}
\author{Yu-Xiao Liu}
\affiliation{Lanzhou Center for Theoretical Physics, Key Laboratory of Theoretical Physics of Gansu Province, School of Physical Science and Technology, Lanzhou University, Lanzhou 730000, China}
\affiliation{Institute of Theoretical Physics \& Research Center of Gravitation, Lanzhou University, Lanzhou 730000, China}

\begin{abstract}
Extracting gravitational wave (GW) signals from individual Galactic binaries (GBs) against their self-generated confusion noise is a key data analysis challenge for  space-borne detectors operating in the $\approx 0.1$~mHz to $\approx 10$~mHz range. Given the likely prospect that there will be multiple such detectors, namely LISA, Taiji, and Tianqin, with overlapping operational periods in the next decade, it is important to examine the extent to which the joint analysis of their data can benefit GB resolution and parameter estimation. To investigate this,  we use realistic simulated LISA and Taiji data containing  the set of $30\times 10^6$ GBs used in the first LISA data challenge (Radler), and an iterative source extraction method called GBSIEVER introduced in an earlier work. We find that a coherent network analysis of LISA-Taiji data boosts the number of confirmed sources by $\approx 75\%$ over that from a single detector. The residual after subtracting out the reported sources from the data of any one of the detectors is much closer to the confusion noise expected from an ideal, but infeasible, multisource resolution method that perfectly removes all sources above a given signal-to-noise ratio threshold. While parameter estimation for sources common to both the single detector and network improves broadly in line with the enhanced signal to noise ratio of GW sources in the latter, deviation from the scaling of error variance predicted by Fisher information analysis is observed for a subset of the parameters. 
\end{abstract}
\maketitle

\section{Introduction}
Gravitational wave (GW) astronomy is now well-established in the frequency band $\approx [10, 1000]$~Hz using ground-based interferometric detectors. Starting with the discovery in 2015 of GW150914~\cite{PhysRevLett.116.061102}, a binary black hole (BBH) merger signal, by the twin LIGO~\cite{aasi2015advanced} detectors, the number of confirmed GW signals has grown to $90$ over three observing runs~\cite{2019GWTC-1,2021GWTC-2,abbott2021gwtc} of the LIGO and Virgo~\cite{acernese2014advanced} detectors. While the majority of signals are from BBH mergers, there is also a double-neutron star~\cite{PhysRevLett.119.161101} and two neutron star-black hole signals~\cite{Abbott_2021} in the haul so far.

The success of GW astronomy with ground-based detectors has further spurred the drive to open up the gravitational wave window at lower frequencies. Pulsar Timing Arrays (PTAs) are already being used for the $\approx [10^{-9}, 10^{-7}]$~Hz and a tantalizing signal, not yet confirmed to be from GWs, has emerged~\cite{NANOGrav:2020bcs,goncharov2021evidence,chen2021common} in the most recent datasets collected by the NANOGrav~\cite{mclaughlin2013north}, PPTA~\cite{manchester2013parkes} and EPTA~\cite{desvignes2016high} collaborations. The $\approx [10^{-4},1]$~Hz band is the target for the space-borne LISA (Laser Interferometer Space Antenna)~\cite{amaro2017laser} detector, scheduled for launch in 2037. Targeting a similar frequency band, intensive work is in progress on the design of the Taiji~\cite{ruan2020taiji} and Tianqin~\cite{luo2016tianqin} missions, both of which have expected launch dates in the middle of the next decade. 

While both LISA and Taiji will have heliocentric orbits, Tianqin is planned to be geocentric. In other respects, however, these detectors share the common feature of having three spacecrafts in a triangular formation that sense GW-induced distance distortion using inter-spacecraft laser links. The nominal distance between the spacecrafts will be $\approx 2.5\times 10^6$~km for LISA and $\approx 3\times 10^6$~km for Taiji, while Tianqin will have shorter arm lengths of $\approx \sqrt{3}\times 10^5\ \mathrm{km}$. The normals to the planes of both the LISA and Taiji constellations will have an approximate tilt of $60^\circ$ from the normal to the ecliptic and rotate once around the latter in a year. For Tianqin, the normal to the spacecraft plane will point towards a fixed sky location, namely that of the double white dwarf RX~J0806.3+1527, at all times.

All planned space-based GW detectors will employ the technique of time-delay interferometry (TDI)~\cite{tinto2021time}, in which the laser frequency Doppler readouts from different arms are combined linearly after introducing time delays, to substantially reduce laser frequency noise. In this paper we consider the so-called $A$, $E$, and $T$ TDI combinations
that have mutually independent noise.

The operational frequency bands of LISA, Taiji, and Tianqin are expected to be rich in astrophysical sources~\cite{Ni:2016wcv,barausse2020prospects}. These include a large population of compact object Galactic binaries (GBs)~\cite{Nissanke:2012eh}, comprised mostly of white dwarfs and reaching $O(10^7)$ in number, tens to hundreds~\cite{baker2019laser} of Extreme Mass Ratio Inspirals (EMRIs) -- a system  containing a massive black hole in the $10^4$ to $10^7\ M_\odot$ range orbited by a stellar-mass range compact object~\cite{PhysRevD.95.103012}, and a handful of Massive black hole binaries (MBHB) containing comparable  $\approx 10^4$ to $10^{6}\ M_\odot$ components~\cite{10.1093/mnras/stz3102}. Due to the generically broad antenna patterns of interferometric detectors and longer lifetimes of GW sources at lower frequencies,  the data from these detectors will be simultaneously occupied by signals from a large number and variety of sources. The multisource resolution problem of disentangling these signals from each other presents a major data analysis challenge that has motivated a number of different approaches and algorithms. 

To provide a benchmark for the development and comparison of data analysis methods, the LISA community has organized several mock LISA data challenges (MLDCs)~\cite{Arnaud:2006gm,arnaud2007overview,Babak2008mock,MockLISADataChallengeTaskForce:2009wir}. The most recent in this series are the LISA data challenges (LDCs)~\cite{baghi2022lisa}. The first LDC (LDC1 or Radler), which is the challenge considered in this paper, consists of several subchallenges among which subchallenge 4 (LDC1-4) focuses purely on GB multisource resolution. LDC1-4 provides single realizations of TDI combinations, each containing Gaussian stationary instrumental noise added to the GW signals from $30\times 10^6$ GBs. A catalog of the parameters of the GBs is also provided, allowing the performance of a multisource resolution method to be quantified rigorously.  

Investigations carried out with the mock data challenges~\cite{MockLISADataChallengeTaskForce:2007iof,MockLISADataChallengeTaskForce:2009wir} indicate that, over an observation period of $\simeq 2$~yr with a single detector, it would be possible to confidently resolve $O(10^4)$ individual GBs in the $\approx [0.1, 15]$~mHz band~\cite{PhysRevD.81.063008,PhysRevD.84.063009,Zhang:2021htc,lu2022implementation}. The remaining GBs will blend together to form a stochastic signal -- the GB background -- that will likely dominate over instrumental noise below $\approx 3$~mHz. Since the GB background will be the main factor limiting the sensitivity of searches for all other GW sources in this band, it is important to develop methods that can extract the maximum number of resolvable GBs without grossly overfitting the data.

Given that, until recently,  LISA was the only anticipated mission, much of the literature on data analysis for space-based detectors has focused on  a single detector. Some notable exceptions are the works  that have analyzed a mission concept called the Big Bang Observer (BBO)~\cite{phinney2004big}, proposed as a successor to LISA. BBO is envisioned to contain $4$ LISA-like detectors but with shorter arm lengths of $\approx 5\times 10^4$~km. Three of the detectors will be spaced $120^\circ$ apart in heliocentric orbits while the fourth will be coincident in location with one of them. In addition to multiple detectors, enhancements such as higher laser power are assumed in order to reach the overall sensitivity expected to reveal the inflationary GW stochastic background. 

While BBO remains a conceptual mission,   the scenario of multiple space-based detectors does not appear to be a pipe dream any longer given the advent of Taiji and Tianqin,  which will likely overlap in operation with LISA. As such, the question of what new science capabilities will be unlocked by the existence of a network of space-based detectors has become a timely one. Several recent works  have approached this question using the computationally inexpensive Fisher information formalism~\cite{PhysRevD.102.024089,PhysRevD.102.063021,PhysRevD.103.103013,Wang:2021uih,Zhang:2021kkh,ruan2021lisa,PhysRevD.105.064055} in the context of single sources. It has been shown that considerable benefits can be derived by combining the data from multiple space-based detectors. However, the analysis of single sources using the Fisher information approach, which only provides an asymptotic lower bound on parameter estimation errors, does not provide a realistic estimate of performance for the full multisource resolution problem. For the latter, there is no substitute for the analysis of realistic mock data with an actual, not ideal, data analysis pipeline.

In this paper, we report an analysis of mock data from a network of two space-based detectors and provide realistic estimates of the detection and estimation performance on the GB multisource resolution problem. The mock data is generated using the same set of GBs as in LDC1-4 under the assumption, which is adequate for a first study and simplifies mock data generation, that Taiji is identical in design to LISA, and that one detector trails the Earth  while the other leads it by $20^\circ$. The mock data is analyzed using the latest version of the GB resolution pipeline, \fitalgoname (Galactic Binary Separation by Iterative Extraction and Validation using Extended Range), introduced in~\cite{Zhang:2021htc}, that has been generalized to handle TDI data from multiple detectors. 

As in the earlier paper~\cite{Zhang:2021htc}, referred to as \firstpaper from here on, it is convenient to define the following  types of GB sources when discussing our results. (i) {\em True}:  sources in the LDC1-4 catalog. (ii) {\em Reported}: the final list of estimated sources returned by \fitalgoname. (iii) {\em Identified}: The initial list of sources produced by the single source search step in \fitalgoname before various cuts are imposed to construct the set of reported sources. (iv) {\em Confirmed}: The reported sources that match true sources as determined by a prescribed metric for association. The fraction of confirmed sources in the set of reported ones is called the {\em detection rate}. 

Our key result is that, at comparable detection rates, a LISA-Taiji network is capable of resolving $\approx 75\%$ more confirmed binaries than either of the detectors alone. In addition, we find that the residual left after subtracting the reported signals from the data tends to have fewer cases of missed strong sources compared to the single detector case. The power spectral density (PSD) of the residual reaches the theoretical lower bound on confusion noise, obtained under the assumption of a perfect but infeasible multisource resolution method for a single detector~\cite{Karnesis:2021tsh}, without significant overfitting. Thus, not only can a detector network probe deeper into the GB population for resolvable sources but the lowered GB background will enhance the detectability of non-GB sources. While parameter estimation errors are reduced with the LISA-Taiji network for sources that are common with the LISA-only analysis, the scaling of error variances with signal strength that is expected from a Fisher information based analysis is not observed for all the parameters.

The rest of the paper is organized as follows. The data used in this paper are described in Sec.~\ref{sec:data}. The baseline single-source detection and parameter estimation method for a detector network is described in Sec.~\ref{sec:snglsrcstat}. A self-contained but brief overview of \fitalgoname is described in Sec.~\ref{sec:method}. This is followed by the results of our analysis in Sec.~\ref{sec:results}. Our conclusions and prospects for future work are presented in Sec.~\ref{sec:conclusions}.

\section{Data description}
\label{sec:data}
Space-based GW detectors will use TDI to strongly suppress laser frequency noise. TDI is implemented by linearly combining readouts of frequency shifts between the incoming and outgoing light along each arm after introducing known time delays~\cite{tinto2021time}. There are a variety of combinations and levels of approximations, called TDI generations, that have been proposed. We use the so-called $A$ and $E$ combinations, which have mutually independent instrumental noise, corresponding to the first generation of TDIs used in LDC1~\cite{babak2020lisa}. (The $T$ combination is dropped because the GW signal in it is highly attenuated.)

To obtain the TDI GW signals, we start with the two polarizations in the transverse traceless (TT) gauge of a plane GW incident at the Solar System barycenter (SSB) origin. For a GB, the polarizations are well-modeled in the SSB frame as linear chirps, 
\begin{align}
h_+(t) & = \gbamp \left(1 + \cos^2\iota\right) \cos \Phi(t)\;,\label{eq:h_plus}\\
h_\times(t) & = -2 \gbamp \cos\iota \sin \Phi(t)\;,\label{eq:h_cross}\\
\Phi(t) & = \phi_0 + 2\pi f t + \pi \dot{f} t^2\label{eq:phase_linchrp}\;,
\end{align}
where $\gbamp$ is the amplitude of the wave, $\iota$ is the inclination angle between the GB orbital angular momentum and the line of sight from the SSB origin to the GB, $\phi_0$ is the initial phase, $f$ is the frequency at the start of observations, and $\dot{f}$ is the secular frequency drift. We use the formalism in~\cite{PhysRevD.81.063008} to generate the TDI GW signals for GB sources.

The TDI time series $\overline{y}^I_D$ for combination $I$ and detector $D$ is given by
\begin{eqnarray}
\overline{y}^I_D & = & \sum_{k=1}^{N_s}\overline{s}^I_D(\theta_k) + \overline{n}^I_D\;, 
\label{eq:data_model}
\end{eqnarray}
where $\overline{x}\in \mathbb{R}^N$ denotes a row vector,  $\overline{s}^I_D(\theta)$ denotes a single GB signal corresponding to source parameters $\theta$, $\overline{n}^I_D$ is a realization of the instrumental noise, and $N_s$ is the number of GBs.  For a GB,  $\theta$ consists of $\{\mathcal{A},\phi_0,\iota,\psi,\lambda,\beta,f,\dot{f}\}$, where $\mathcal{A}$, $\phi_0$, $\iota$, $f$, and $\dot{f}$ were defined following Eq.~\ref{eq:h_plus} to Eq.~\ref{eq:phase_linchrp}),  $\psi$ is the polarization angle defining the orientation of the binary orbit projected on the sky, and the longitude and the latitude of the source in the SSB frame are denoted  by $\lambda$ and $\beta$, respectively. In the rest of the paper,  $\theta$ will serve as a stand in for a GB source itself where convenient. Following LDC1-4 we set the sampling frequency for the uniformly sampled time series $\overline{y}^I_D$, for all $I$ and $D$, to be $f_s = 1/15$~Hz, and the number of samples to $N=4194304$  corresponding to an observation period  $T_{\rm{obs}}\approx2$~yr.

For the generation of mock TDI data, the LISA community has adopted a standard code called LISACode~\cite{Petiteau:2008zz} for generating instrumental noise and a code called FastGB for GB signals. (The latter is based on the formalism in~\cite{PhysRevD.76.083006} and is included in the software distribution associated with LDC data~\cite{LDC1-website}.) The simulated LDC1-4 data contains the TDI signals from $30$ million GBs, with GW signal frequencies in  $\approx[0.1,15]$~mHz. In the astrophysically realistic model used for the GB population~\cite{nelemans2013galactic,korol2020populations}, the density of GBs drops with increasing frequency with most of the sources above $4$~mHz being individually resolvable. The instrumental noise realization is drawn from a Gaussian, stationary, stochastic process. The PSD of the noise is denoted by $S_{D}^I(\nu)$ at Fourier frequency $\nu$. In LISACode, the PSD of the instrumental noise in the A and E combinations are identical and derived from the design sensitivity of LISA. 

We generate simulated Taiji TDI data by shifting the centroid of the orbit $40^\circ$ ahead of LISA and using the same software suite and GB catalog as LDC1-4. Statistical independence of the instrumental noise in the two detectors is ensured by using different seeds for the corresponding pseudo-random sequences in LISACode. We keep the arm length of Taiji the same as that of LISA. Besides allowing LISACode to be used with minimal changes at the code level, this leads to a simpler and more direct comparison, as befits a first study, of the performance gained in switching from a single detector to a network. To acknowledge this simplification, the simulated Taiji data is called \Taijimath in the remainder of the paper. Extending our analysis to a realistic Taiji would be quite straightforward when its own data simulation framework is made available in the public domain.

The GW signal in each TDI combination has amplitude and frequency modulations  arising from the time-varying antenna response due to the rotation of the spacecraft constellation and the Doppler effect due to its orbital motion, respectively.  Consequently, the spectrum of the TDI GB signal is broadened over a frequency range that is $\approx 10^4\times f$ times larger than the separation of Fourier frequencies for a $\approx 2$~yr observation period. The broadening increases the overlap of spectra from multiple GBs, with the number of overlaps increasing at lower frequencies and signal strengths. This precludes the differentiation of the signals using a simple Fourier transform and makes multisource resolution an extremely challenging data analysis problem. 

\section{Single source estimation}
\label{sec:snglsrcstat}
\fitalgoname implements an iterative scheme in which the parameters of only a single source are estimated at a time and the corresponding signal is subtracted from the data. The parameter estimation step uses maximum likelihood estimation (MLE) where the log-likelihood function is constructed under the assumption that the data contains a single source added to Gaussian, stationary noise. The maximization of the log-likelihood in the case of a GB can be carried out analytically over a subset of the parameters, leaving behind a function, commonly called the \fstat in the GW literature~\cite{Jaranowski:1998qm}, that needs to be numerically maximized over the remaining ones.  In \fitalgoname, the latter maximization is carried out using particle swarm optimization (PSO)~\cite{eberhart1995particle}. 

The principal difference between P1 and the present paper is the shift in the mathematical formalism implemented in \fitalgoname, which is a straightforward generalization from a single detector to, as described in this section, a network of detectors.

\subsection{Network \fstat}
\label{sec:fstat}
The GW signal from a single GB in combination $I$ of detector $D$ can be expressed, schematically, as
\begin{align}
    \overline{s}^I_D(\theta) & = \overline{a}\,{\bf X}^I_D(\intpar)\;,
\end{align}
where the {\em extrinsic} parameters $\overline{a}=(a_1, a_2, a_3, a_4)\in \mathbb{R}^4$ are  obtained by reparametrizing $\gbamp$, $\phi_0$, $\psi$, and $\iota$.  ${\bf X}^I_D(\intpar)$ is a $4$-by-$N$ matrix of {\em template} waveforms that depend on the {\em intrinsic} parameter set $\intpar = \{\lambda, \beta, f, \dot{f}\}$. 
[In the GW literature, the terms intrinsic and extrinsic generally refer to sets of 
parameters over which the log-likelihood is maximized using numerical and analytical (or computationally efficient) methods, respectively.]

The estimator, $\widehat{\theta}_M$, of the parameters $\theta$ of a single source in  iteration $M\geq 1$ is given by,
\begin{align}
    \widehat{\theta}_M & = \argmin_{\theta} 
    \sum_{I\in\mathcal{I},D\in\mathcal{D}}
    \left(\| \overline{y}^I_{D,M} -  \overline{s}^I_D(\theta)\|^I_D\right)^2\;,
    \label{eq:mletheta}
\end{align}
where $\mathcal{D}$ is the set of detectors, $\mathcal{I}$ is the set of TDI combinations (with mutually independent noise) per detector, $\overline{y}^I_{D,M} = \overline{y}^I_{D,M-1}-\overline{s}^I_D(\widehat{\theta}_{M-1})$ and $\overline{y}^I_{D,1} = \overline{y}^I_D$ as defined in Eq.~\ref{eq:data_model}. (We drop the iteration index and the source parameter, where convenient, in the expressions below for simplicity of notation.) Here, $\|.\|^I_D$ is the norm induced by the noise-weighted inner product,
\begin{align}
    \langle \overline{x},\overline{z}\rangle^I_D & = \frac{1}{Nf_s}(\widetilde{x}./\overline{S}_{D}^I)\widetilde{z}^\dagger\;,
    \label{eq:innprod}
\end{align}
where $f_s$ is the sampling frequency, $\widetilde{x}^T  = {\bf F}\overline{x}^T$ is the discrete Fourier transform (DFT) of $\overline{x}$ [defined by $F_{kl} = \exp(-2\pi i k l/N)$], `$./$' denotes element-wise division, and $\overline{S}_{D}^I$ is the sequence of $S^I_D(\nu)$ samples at the DFT frequencies. A note on implementation: It is computationally efficient to confine searches for single sources to narrow frequency bands (see Sec.~\ref{sec:search_bands}) and analyze different bands  in parallel on a multi-processor machine. The PSD $S_D^I(\nu)$ in a sufficiently narrow search band is well approximated by a constant, turning the inner product in Eq.~\ref{eq:innprod} to an ordinary  Euclidean one between bandpassed time series (see Sec.~\ref{sec:undersampling}).  

Let $\overline{A}(i)$ denote  row $i$ and $A(i,j)$ the element in row $i$ and column $j$
of a matrix ${\bf A}$. Let ${\bf U}^I_D$ denote the column vector with the $i$th element
\begin{align}
    U^I_D(i) &= \langle \overline{y}^I_D, \overline{X}^I_D(i)\rangle^I_D\;,
    \label{eq:umat}
\end{align}
and ${\bf W}^I_D$ denote the matrix with
\begin{align}
    W^I_D(i,j) &= \langle \overline{X}^I_D(i),
                            \overline{X}^I_D(j)\rangle^I_D\;.
    \label{eq:wmat}
\end{align}
Then, the minimization problem in Eq.~\ref{eq:mletheta} can be recast as a maximization,
\begin{align}
    \widehat{\theta} & =  \argmax_\theta \sum_{I\in\mathcal{I},D\in\mathcal{D}}\left(
            \overline{a}{\bf U}^I_D - \frac{1}{2}\overline{a}{\bf W}^I_D\overline{a}^T
    \right)\nonumber\\
    & = \argmax_\theta \left( \overline{a}{\bf U} - \frac{1}{2}\overline{a}{\bf W}\overline{a}^T\right)\;,
\end{align}
where ${\bf W} = \sum_{I\in\mathcal{I},D\in\mathcal{D}} {\bf W}^I_D$ and ${\bf U} = \sum_{I\in\mathcal{I},D\in\mathcal{D}} {\bf U}^I_D$. For fixed $\kappa$, the maximization over $\overline{a}$ is trivial,
\begin{align}
    \widehat{a}^T & = {\bf W}^{-1}{\bf U}\;.
    \label{eq:amp_est}
\end{align}
The estimator of $\kappa$ is then given by 
\begin{align}
    \widehat{\kappa} & =  \argmax_\kappa \mathcal{F}(\kappa)\;,
    \label{eq:int_est}
\end{align}
where
\begin{align}
       \mathcal{F}(\kappa) & = {\bf U}^T {\bf W}^{-1}{\bf U}\;
       \label{eq:Fstat}
\end{align}
is widely known in the GW literature as the \fstat. The estimated extrinsic parameters $\widehat{a}$ are obtained by substituting $\widehat{\intpar}$ in Eq.~\ref{eq:amp_est}. 

\subsection{Signal-to-noise ratio and association metric}
\label{sec:snr}
It is convenient  to use the signal-to-noise ratio (SNR) defined below to represent the overall strength of a signal relative to instrumental noise.
\begin{align}
    {\rm SNR}^2 & = \sum_{I\in\mathcal{I},D\in\mathcal{D}}\left(\|\overline{s}^I_D(\theta)\|^I_D\right)^2\;,
    \label{eq:snr}
\end{align}
where the norm is defined under the constant PSD approximation described earlier. For the LDC1-4 and \Taijimath $A$ and $E$ TDI combinations, $S_D^I(\nu)$ is independent of both $D$ and $I$, thereby appearing as a constant overall factor in Eq.~\ref{eq:mletheta} for a given search band. As such, the value of $S_D^I(\nu)$ used for a search band affects the conversion of the overall amplitude $\mathcal{A}$ to SNR but does not affect $\widehat{\theta}$. (The error made in the estimation of $\theta$, on the other hand, does depend on the true $S_D^I(\nu)$ as it governs the level of noise in $\overline{y}_D^I$.)

The independence of $S_D^I(\nu)$ from $D$ also implies that the SNR of a GB signal is amplified by a factor of $\leq \sqrt2$ in the two detector network relative to a single detector. For a multi-year long observation period, the signal from a source in both LISA and Taiji-mod undergoes the same degree of amplitude and frequency modulations, which makes the summands in Eq.~\ref{eq:snr} comparable. Hence, the increase in SNR is close to $\sqrt{2}$ for all sources. While one expects both source identification and parameter estimation to improve for an individual source due to the increase in SNR, this is not sufficient in itself to explain the impact of a detector network on multisource resolution. The details of how the sources mutually interfere with each other also play an important role and need to be taken into account.

In analyzing the output of a multisource resolution method, a metric is required to quantify the degree of association  between a given pair of sources. As in P1, and across the LISA mock data challenges in general, it is convenient for this purpose to use the correlation coefficient, $R(\theta,\theta^\prime)$, between the TDI signals corresponding to a given pair of sources, $\theta$ and $\theta^\prime$, as follows.
\begin{align}
    R(\theta,\theta^\prime) & = \frac{C(\theta,\theta^\prime)}{\left[C(\theta,\theta)C(\theta^\prime,\theta^\prime)\right]^{1/2}}\;,\\
    C(\theta,\theta^\prime) & = \sum_{I\in\mathcal{I},D\in\mathcal{D}}\langle \overline{s}^I_D(\theta),\overline{s}^I_D(\theta^\prime)\rangle^I_D\;.
\end{align}
Since the correlation coefficient only measures similarity between the shapes of the signals, it needs to be supplemented by an SNR-based criterion to account for the closeness of sources in their overall amplitudes. For a pair of reported and true sources, denoted by $\widehat{\theta}$ and $\theta$, respectively, \fitalgoname uses the scheme laid out in MLDC-3~\cite{MockLISADataChallengeTaskForce:2009wir} wherein such a pair is eligible for association only when  $\theta$ has (a) an  ${\rm SNR} \geq 3$, (b) a frequency within $6$ DFT frequencies of $\widehat{f}$, and (c) the lowest distance, defined as $\left[\sum_{I,D}(\|\overline{s}^I_D(\widehat{\theta})-\overline{s}^I_D(\theta)\|^I_D)^2\right]^{1/2}$, from $\widehat{\theta}$. Given an eligible association, the status of a reported source is elevated to confirmed only if $R(\theta,\widehat{\theta}) \geq 0.9$.

The test outlined above does not prevent multiple reported sources from being associated with the same true source.  We follow~\cite{PhysRevD.81.063008} in handling such an ambiguous case by retaining only the reported source with the highest $R(\theta,\widehat{\theta})$ value provided $R(\theta,\widehat{\theta}) \geq 0.9$. This approach is a conservative one since it lowers the detection rate by reducing the count of confirmed sources but not that of reported ones. We actually did not find any such case in the analysis of the network data while, for single detector search, there is only one instance. Hence, the effect on the detection rate is negligible in practice for the data considered in this paper.

\section{Overview of GBSIEVER and its settings}
\label{sec:method}
The principal algorithmic components of \fitalgoname and their corresponding settings are reviewed in this section. The description is self-contained but brief since further details are available in \firstpaper. We do not provide an extensive review of PSO here, which is used for the global optimization task in Eq.~\ref{eq:int_est}, since it is a widely known algorithm. A pedagogical introduction to PSO and its application to MLE can be found in~\cite{mohanty2018swarm}. The specific variant of PSO used in \fitalgoname is discussed in \firstpaper and its parameter settings are described in detail in Sec. III of~\cite{normandin2018particle}.

\subsection{Narrow frequency band search and edge effects}
\label{sec:search_bands}
As mentioned earlier, single source searches in \fitalgoname are performed in parallel over narrow frequency bands. The width of each search band is set at $0.02$~mHz but only the sources found in the central $0.01$~mHz, called the {\em acceptance zone}, are admitted into the set of identified sources. This restriction is imposed to counter the excessive occurrence of spurious identified sources near the band edges due to the spread of signal power from strong sources across the search bands. The frequency band limits are imposed by applying a Tukey window to the DFT of a TDI time series. The central flat part of the window is kept slightly wider, at $0.015$~mHz, than the acceptance zone. The sources discarded in a given search band are not lost but recovered in adjacent ones because they are overlapped to make all the acceptance zones contiguous.

\subsection{Undersampling}
\label{sec:undersampling}
After applying the Tukey window for a given search band, an inverse DFT brings the bandpassed TDI data back into the time domain. Here, a key step that follows is the {\em undersampling}~\cite{donoho2010precise} of the time series, which drastically reduces the number of samples without any loss of information. Undersampling is a clever technique that exploits the aliasing error caused by sampling below the Nyquist rate to move the information content in bandpassed data to low frequencies. Recalling that the \fstat in a given search band is evaluated under the white noise approximation, allowing the inner product in Eq.~\ref{eq:innprod} to be evaluated in the time domain, the reduction in the number of samples by undersampling leads to a much faster computation of the \fstat.

\subsection{Termination rule}
The single-source estimation and subtraction iterations in a search band are terminated when (i) $300$ sources have been identified, or (ii) the estimated source SNRs in $5$ consecutive iterations fall below $7.0$. We note that in P1, the maximum number of iterations was set at $200$ but this was never reached for any of the search bands as criterion (ii) was satisfied first. Hence, the increase in the maximum number of iterations here does not affect the single detector results. However, the larger number of iterations is required in the case of the detector network because, for the same SNR threshold, sources have a lower amplitude, hence are more numerous, relative to a single detector.

\subsection{Cross-validation}
\label{sec:cross-validation}
Besides PSO, a key feature of \fitalgoname that differentiates it from other multisource resolution methods that have been applied to mock LISA data is the step called {\em cross-validation}. In this step, the entire single source search is rerun on the data with all settings held the same except for the search range used for the secular frequency drift $\dot{f}$. Thus, we obtain two sets of identified sources from the two runs. In one of the runs, called the {\em primary}, the $\dot{f}$ range is governed by the expected astrophysical one, while in the other (called {\em secondary}) it is set much wider. Finally, for each identified source $\widehat{\theta}_{1,i}$ in the primary run, we compute
\begin{align}
    R_{\rm ee}(\widehat{\theta}_{1,i}) & = \max_j R(\widehat{\theta}_{1,i},\widehat{\theta}_{2,j})\;,
\end{align}
with $\widehat{\theta}_{2,j}$ belonging to the set of identified sources from the secondary run. The identified source $\widehat{\theta}_{1,i}$ is elevated to the status of a reported source only if $R_{\rm ee}(\widehat{\theta}_{1,i})$ crosses a preset threshold. As demonstrated in P1, cross-validation is highly effective in eliminating spurious identified sources. This is because they are much less likely to recur in the two runs unlike the identified sources that are associated with true sources.

We impose different $R_{\rm ee}$ thresholds in different blocks that tile the SNR and frequency plane contiguously. This addresses the expectation that the incidence of spurious sources is not constant across this plane: In a given frequency interval, the fraction of spurious sources increases at lower SNRs, requiring a higher $R_{\rm ee}$ cutoff to weed them out, while the fall-off in the density of GBs with increasing frequency leads to less confusion and fewer spurious sources, requiring a less stringent $R_{\rm ee}$ cutoff.
 
The combination of blocks in the SNR-frequency plane and the associated $R_{\rm ee}$ thresholds are part of the user-defined settings in \fitalgoname. In P1, we carried out a comparative analysis of different combinations. Here, we simply  run \fitalgoname for the particular combination below that was used in P1 to obtain the principal results. 
\begin{align}
    R_{\rm ee} & = \left\{ \begin{array}{cc}
                      0.9, & \nu\in [0,3]~{\rm mHz}, {\rm SNR} \leq 25\,\\
                      0.5, & \nu\in [0,3]~{\rm mHz}, {\rm SNR} > 25
                             \end{array}
              \right.\\
    R_{\rm ee} &= \left\{ \begin{array}{cc}
                      0.9, & \nu\in [3,4]~{\rm mHz}, {\rm SNR} \leq 20\,\\
                      0.5, & \nu\in [3,4]~{\rm mHz}, {\rm SNR} > 20
                             \end{array}
              \right.
\end{align}
Cross-validation is not required for $\nu > 4$~mHz because sources can be well resolved without generating a significant number of spurious ones.

\section{Results}
\label{sec:results}
The presentation of results in this section is organized as follows. Sec.~\ref{sec:resolution_results} contains results on source resolution. An investigation of the residual is presented in Sec.~\ref{sec:residuals}. Sec.~\ref{sec:paramest_results} contains results on GB parameter estimation.

\subsection{Source resolution performance}
\label{sec:resolution_results}
To quantify the performance of \fitalgoname in GB resolution, we use the detection rate, defined as the fraction of reported sources that are elevated to the status of confirmed sources using the test of association described in Sec.~\ref{sec:snr}.

Table~\ref{tab:snr_ree_LISA_Taiji} provides a summary of the output from \fitalgoname for the LISA and Taiji network using the LDC1-4 and Taiji-mod data. For comparison, the results from the analysis of LDC1-4 data alone are reported in Table~\ref{tab:snr_ree_LISA}. The primary observation is that the number of confirmed GBs is boosted from $10,388$ for LISA to $18,151$ for the LISA-Taiji network, a remarkable increase of $74.73\%$. Similarly the number of confirmed sources in corresponding frequency ranges are also significantly higher for the LISA-Taiji network. For example, an increase of $\approx 5,500$ in the frequency range, $\nu\in [0,3]$~mHz, was achieved.

\begin{table*}
    \centering
    \begin{tabular}{|c|c|c|c|c|c|c|c|c|c|c|}
    \hline
    \cline{2-11}
        &$\nu$~mHz &  SNR & $\nu$~mHz &  SNR & $\nu$~mHz &  SNR & $\nu$~mHz &  SNR &  $\nu$~mHz &  SNR  \\
    \cline{2-11}
        &$[0,3]$ & $[0,25]$ & $[0,3]$ & $[25,\infty]$ & $[3,4]$ & $[0,20]$ 
        &$[3,4]$ & $[20,\infty]$ & $[4,15]$ & $[10,\infty]$ \\
    \hline
   $R_{\rm ee}$ & \multicolumn{2}{|c|}{$0.9$} & \multicolumn{2}{|c|}{$0.5$} &
               \multicolumn{2}{|c|}{$0.9$} & \multicolumn{2}{|c|}{$0.5$} & \multicolumn{2}{|c|}{$-1$}\\
       \hline
   Identified   & \multicolumn{2}{|c|}{$33144$} & \multicolumn{2}{|c|}{$3941$} &
               \multicolumn{2}{|c|}{$3461$} & \multicolumn{2}{|c|}{$2528$} & \multicolumn{2}{|c|}{$4687$}\\
   \hline
   Reported     & \multicolumn{2}{|c|}{$8420$} & \multicolumn{2}{|c|}{$3920$} &
               \multicolumn{2}{|c|}{$2440$} & \multicolumn{2}{|c|}{$2526$} & \multicolumn{2}{|c|}{$4687$}\\
   \hline
   Detection rate  & \multicolumn{2}{|c|}{$66.05\%$} & \multicolumn{2}{|c|}{$91.96\%$} &
               \multicolumn{2}{|c|}{$83.73\%$} & \multicolumn{2}{|c|}{$95.25\%$} & \multicolumn{2}{|c|}{$96.78\%$}\\
   \hline
   Lowest SNR (confirmed)  & \multicolumn{4}{|c|}{$7.1$} & \multicolumn{4}{|c|}{$6.6$} &
               \multicolumn{2}{|c|}{$10.1$}\\
    \hline
    \multicolumn{1}{|c|}{Total reported} & \multicolumn{10}{|c|}{$21993$}\\
    \hline
    \multicolumn{1}{|c|}{Total confirmed} & \multicolumn{10}{|c|}{$18151$}\\
    \hline
    \multicolumn{1}{|c|}{Detection rate} & \multicolumn{10}{|c|}{$82.53\%$}\\
    \hline
    \end{tabular}
    \caption{ Performance of \fitalgoname for the LISA-Taiji network using  LDC1-4 and Taiji-mod data. The results are organized according to the contiguous blocks in the SNR and frequency plane used for cross-validation. The primary and secondary search ranges for $\dot{f}$ used in cross-validation for $f\leq4$~mHz are $[-10^{-16},  10^{-15}]~\rm{Hz^2}$ and $[-10^{-14}, 10^{-13}]~\rm{Hz^2}$ respectively. For $f\in[4,15]$~mHz, only the primary search is used with the range $[-10^{-14}, 10^{-13}]~\rm{Hz^2}$. A block for which the $R_{\rm ee}$ cut was not used is shown as having $R_{\rm ee}=-1$. \label{tab:snr_ree_LISA_Taiji}}
\end{table*}
\begin{table*}
    \centering
    \begin{tabular}{|c|c|c|c|c|c|c|c|c|c|c|}
    \hline
    \cline{2-11}
        &$\nu$~mHz &  SNR & $\nu$~mHz &  SNR & $\nu$~mHz &  SNR & $\nu$~mHz &  SNR &  $\nu$~mHz &  SNR  \\
    \cline{2-11}
        &$[0,3]$ & $[0,25]$ & $[0,3]$ & $[25,\infty]$ & $[3,4]$ & $[0,20]$ 
        &$[3,4]$ & $[20,\infty]$ & $[4,15]$ & $[10,\infty]$ \\
    \hline
   $R_{\rm ee}$ & \multicolumn{2}{|c|}{$0.9$} & \multicolumn{2}{|c|}{$0.5$} &
               \multicolumn{2}{|c|}{$0.9$} & \multicolumn{2}{|c|}{$0.5$} & \multicolumn{2}{|c|}{$-1$}\\
       \hline
   Identified   & \multicolumn{2}{|c|}{$23231$} & \multicolumn{2}{|c|}{$2106$} &
               \multicolumn{2}{|c|}{$3696$} & \multicolumn{2}{|c|}{$1526$} & \multicolumn{2}{|c|}{$4279$}\\
   \hline
   Reported     & \multicolumn{2}{|c|}{$2767$} & \multicolumn{2}{|c|}{$2073$} &
               \multicolumn{2}{|c|}{$1622$} & \multicolumn{2}{|c|}{$1510$} & \multicolumn{2}{|c|}{$4279$}\\
   \hline
   Detection rate  & \multicolumn{2}{|c|}{$63.61\%$} & \multicolumn{2}{|c|}{$91.27\%$} &
               \multicolumn{2}{|c|}{$80.33\%$} & \multicolumn{2}{|c|}{$92.32\%$} & \multicolumn{2}{|c|}{$94.39\%$}\\
   \hline
   Lowest SNR (confirmed)  & \multicolumn{4}{|c|}{$7.7$} & \multicolumn{4}{|c|}{$6.8$} &
               \multicolumn{2}{|c|}{$10.0$}\\
    \hline
    \multicolumn{1}{|c|}{Total reported} & \multicolumn{10}{|c|}{$12251$}\\
    \hline
    \multicolumn{1}{|c|}{Total confirmed} & \multicolumn{10}{|c|}{$10388$}\\
    \hline
    \multicolumn{1}{|c|}{Detection rate} & \multicolumn{10}{|c|}{$84.79\%$}\\
    \hline
    \end{tabular}
    \caption{Performance of \fitalgoname for the single-detector  LDC1-4 data. The results are organized according to the contiguous blocks in the SNR and frequency plane used for cross-validation. The primary and secondary search ranges for $\dot{f}$ used in cross-validation for $f\leq4$~mHz are $[-10^{-16},  10^{-15}]~\rm{Hz^2}$ and $[-10^{-14}, 10^{-13}]~\rm{Hz^2}$ respectively. For $f\in[4,15]$~mHz, only the primary search is used with the range $[-10^{-14}, 10^{-13}]~\rm{Hz^2}$. A block for which the $R_{\rm ee}$ cut was not used is shown as having $R_{\rm ee}=-1$. The numbers in this table correspond to those in Table~I of \firstpaper for the combination of SNR and $R_{\rm ee}$ cuts called {\em Main}. However, there are small differences (e.g., total reported changed from $12270$ to $12251$ while detection rate went from $84.28\%$ to $84.79\%$) due to the cumulative effect of several small code changes made to \fitalgoname since \firstpaper.}
    \label{tab:snr_ree_LISA}
\end{table*}

The overall detection rate of the LISA-Taiji network is $82.53\%$, which is slightly smaller than the $84.79\%$ for LISA. However, if the first frequency range ($[0,3]$~mHz),  which has the lowest detection rate is ignored, the detection rate for the LISA-Taiji network becomes $93.08\%$ while that for LISA becomes smaller at $90.89\%$. Thus, while the overall detection rate does not show a consistent preference for a single detector or a network, the latter  finds far more reported sources overall, causing a corresponding jump in the number of confirmed sources.

An auxiliary metric that is useful for gauging the performance of a multisource resolution method is the lowest SNR among confirmed sources. This tells us how deep into the source population can we dive to extract resolvable sources. While it would appear from Tables~\ref{tab:snr_ree_LISA_Taiji} and~\ref{tab:snr_ree_LISA} that there is no significant change  in this quantity, it is important to note that the SNRs are computed for different numbers of detectors. In terms of the signal amplitude, $\mathcal{A}$, which is the invariant quantity here, the SNR for a two detector network would be $\approx \sqrt{2}$ times higher than that for a single detector. Conversely, the same SNR implies that the signal in the network has $\mathcal{A}$ that is a factor of $\approx \sqrt{2}$ lower. Hence, the LISA-Taiji network is able to increase the search depth quite substantially.

\subsection{Residuals}
\label{sec:residuals}
\begin{figure*}
    \centering
    \includegraphics[width=\textwidth]{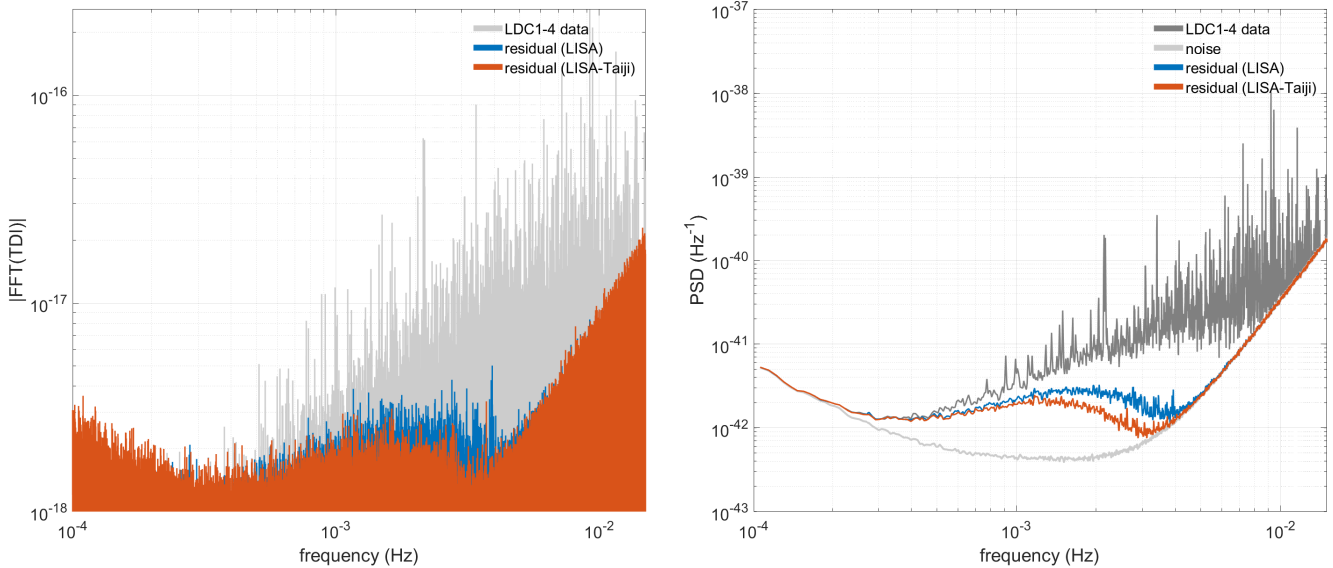}
    \caption{Absolute value of the DFT (left) and PSD (right) of the data and residuals. In both panels, the gray curves correspond to LDC1-4 TDI $A$ data, while the blue and red curves correspond to the residuals obtained by subtracting out the reported sources found from the analysis of single-detector (LISA) and network (LISA-Taiji) data, respectively. The frequency resolution of the PSD is $8.1380\times 10^{-3}$~mHz. For visual clarity, we have imposed a lower cutoff on the Y-axis range in the left panel since there is no useful information below this cutoff.}
    \label{fig:residual}
\end{figure*}

Fig.~\ref{fig:residual} compares the spectral properties of the TDI $A$ data and the residual obtained by subtracting out all the reported sources from it. There is a significant reduction of spectral power in the LISA-Taiji residual relative to the case of a single detector in the $\approx [1, 3]$~mHz band. The residuals begin to converge to the instrumental noise in the $\approx [3, 4]$~mHz band. The Pearson correlation coefficient between the residual and instrumental noise time series in the $A$ combination bandpassed to $[3, 4]$~mHz improves from $0.6636$ for the single detector case to $0.8632$ for the LISA-Taiji network. If one could remove all true sources perfectly from the data, the residual would have a correlation of unity with the instrumental noise. Thus, the significantly improved correlation above shows that the set of reported sources found in this band by the network is more reliable and complete than the one from a single detector.

The LISA-Taiji residual is also less spiky between $3$ to $4$~mHz because a cluster of loud sources that was missed in the single detector analysis was identified and extracted out. In general, defining loud sources as those with a single-detector true ${\rm SNR}\geq 20$, the LISA-Taiji network finds $600$ additional loud sources across the entire frequency range, and the fraction of true loud sources that were not detected falls from $11\%$ to $3\%$. 

The subtraction of reported GB sources from the data is a prerequisite for mounting a search for any other type of GW source. Since the signals from the GBs in the data will superpose to form the GB background noise that will limit the sensitivity of these searches, it is important to understand how close the actual GB background comes to the expected one after the subtraction of reported sources. However, what constitutes the irreducible expected GB background is not a straightforward question. One approach, reviewed in brief below, is the scheme proposed in~\cite{PhysRevD.73.122001} and developed further in~\cite{Karnesis:2021tsh}. 

It is assumed that one has an ideal  method that is capable of estimating the parameters of a source perfectly as long as it has a minimum SNR. The SNR here is defined relative to the floor of the PSD of the data, estimated using  a smoothing method such as running mean or median~\cite{mohanty2003efficient}, not the instrumental noise PSD. For clarity, let us denote it as ${\rm SNR}_{\rm bg}$ and the minimum value as ${\rm SNR}_{\rm bg}^{\rm min}$. The subtraction of sources and the estimation of the PSD floor follow each other iteratively until no sources with ${\rm SNR}_{\rm bg}\geq {\rm SNR}_{\rm bg}^{\rm min}$  are left in the data. The remaining data is then deemed to be devoid of resolvable sources, hence an estimate of the GB background. There are some important caveats to this approach, one being that it is only applicable to data from a simulated population of sources with known parameters, and the other being the unrealizable assumption of an ideal method. The choice of ${\rm SNR}_{\rm bg}^{\rm min}$, which is ad hoc in itself, is also an important factor that governs the final GB background. Nonetheless, it is reasonable to accept the ideal GB background obtained from this approach to be a lower bound on the actual one that can be achieved by a practical multisource resolution method.

\begin{figure}
    \centering
    \includegraphics[scale=0.36]{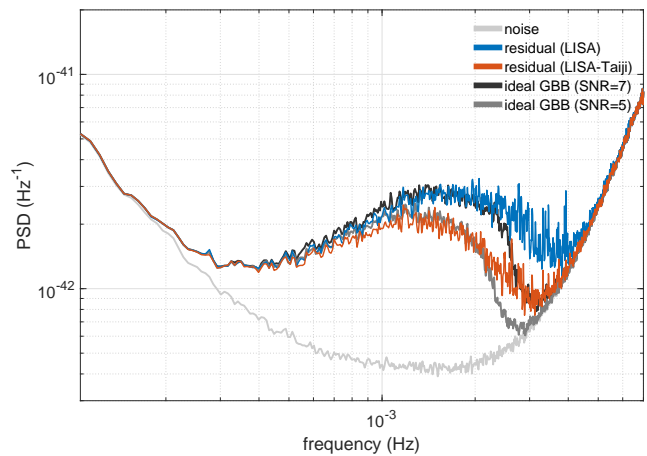}
    \caption{Residuals obtained with the single-detector (blue) and network analysis (red)  compared with the ideal GB background (GBB) for ${\rm SNR}_{\rm bg}^{\rm min} = 5.0$ (dark gray) and $7.0$ (black). The blue and red curves are identical to the corresponding ones in Fig.~\ref{fig:residual}. We experimented with the settings of the algorithm for the ideal GBB such that the black curve matches the one in Fig.~1 of~\cite{Karnesis:2021tsh} as closely as possible. The dark gray curve is from the same settings except for the lower ${\rm SNR}_{\rm bg}^{\rm min}$. }
    \label{fig:ideal_confusion}
\end{figure}

Fig.~\ref{fig:ideal_confusion} compares the LISA-Taiji residual (same as in Fig.~\ref{fig:residual}) with the ideal GB background obtained, as outlined above, for LISA alone using ${\rm SNR}_{\rm bg}^{\rm min}\in \{5.0,7.0\}$.  We see that the LISA-Taiji residual lies well below the single-detector ideal GB background for ${\rm SNR}_{\rm bg}^{\rm min} = 7.0$ and nearly coincides with that for ${\rm SNR}_{\rm bg}^{\rm min} = 5.0$ at frequencies below $\approx 2$~mHz. The latter is an unrealistically optimistic lower bound on the background since one does not expect any algorithm to confidently resolve sources perfectly at this SNR. However, using the LISA-Taiji network will still allow this level of the background to be reached. 

While the LISA-Taiji residual actually dips a little under the ${\rm SNR}_{\rm bg}^{\rm min} = 5.0$ ideal background below $\approx 2$~mHz, this is the effect of overfitting the data caused by the presence of spurious sources at low SNR. Overfitting is an important effect to consider for any multisource resolution method, not just \fitalgoname, since spurious sources are unavoidable. The overfitting of data can result in a loss in SNR for a non-GB source if its spectral power is close in level to the ideal background. As seen in Fig.~\ref{fig:ideal_confusion}, the observed overfitting is quite mild for \fitalgoname running on network data and we do not expect it to be a major issue. However, this discussion also points to the necessity of subjecting all multisource methods to a comparison with the ideal background in order to gauge their tendency to overfit the data. 

\subsection{Parameter estimation performance}
\label{sec:paramest_results}
\begin{figure*}[h]
    \centering
    \includegraphics[width=\textwidth]{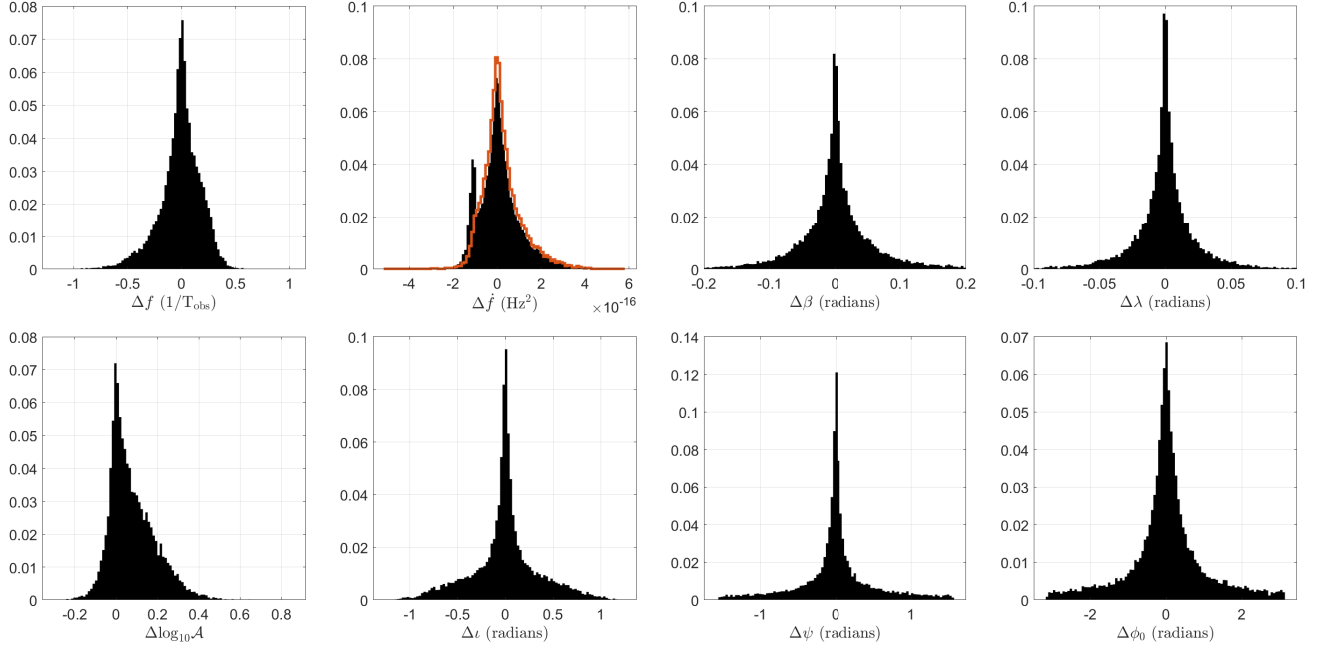}
    \caption{Estimated PDFs (normalized histograms) of the differences in parameter values of confirmed and associated true sources obtained from the LISA-Taiji network. Small fractions of outliers in the distributions of $\Delta\beta$ and $\Delta\lambda$, constituting $2.0\%$ and $1.5\%$ of the full sample, respectively, have been dropped for visual clarity. The secondary peak in the PDF of $\Delta\dot{f}$ is an artifact produced by the estimates for some sources accumulating along the $\dot{f}$ search boundary used in PSO for the primary run. Removing these on-edge estimates, in the range $[-10^{-16}, -0.99\times10^{-16}]$~${\rm Hz}^2$, suppresses the secondary peak as shown by the red curve. }
    \label{fig:network-errors}
\end{figure*}

For assessing the parameter estimation performance of \fitalgoname, we consider the difference between the values of a parameter for a confirmed source and its associated true source. Fig.~\ref{fig:network-errors} shows the estimated probability density function (PDF) of this difference for each GB signal parameter.  Comparing them with the corresponding ones in \firstpaper for the single-detector case, we find no significant differences. [In terms of the actual counts per bin, however, there would be a substantial difference due to the larger number of confirmed sources (c.f., Tables~\ref{tab:snr_ree_LISA_Taiji} and~\ref{tab:snr_ree_LISA}) for the network.]

\begin{figure*}
    \centering
    \includegraphics[width=\textwidth]{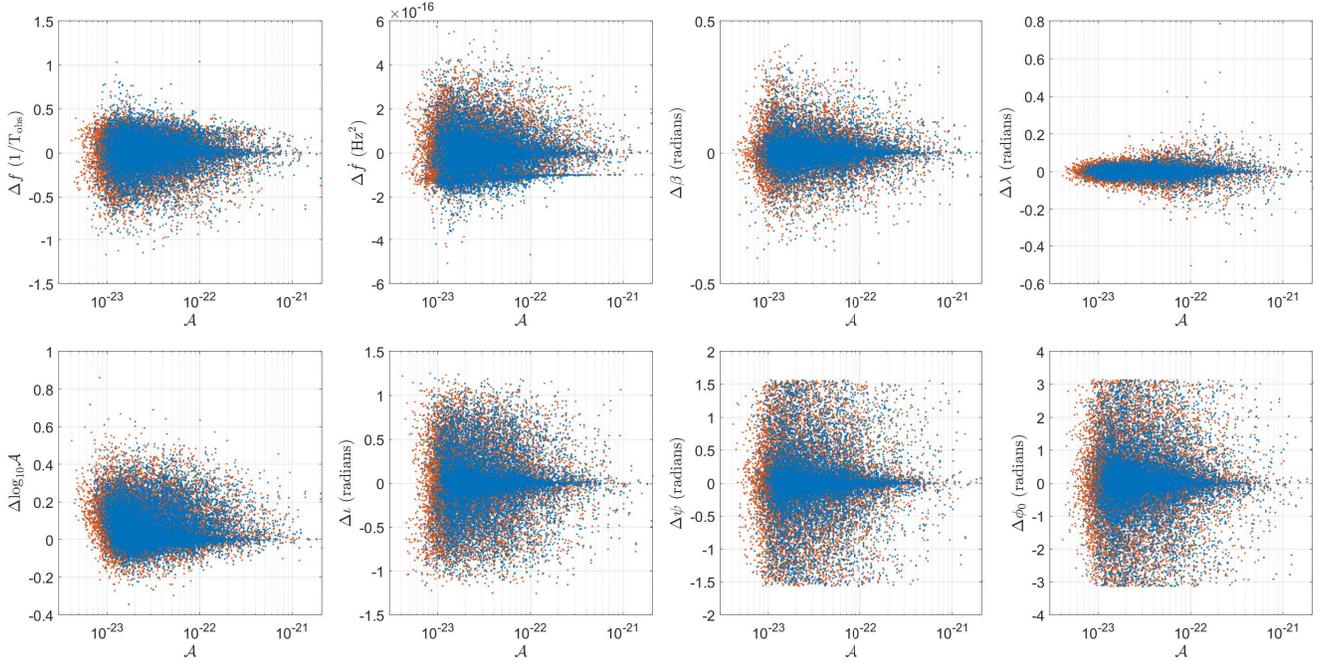}
    \caption{ Differences in the parameter values of confirmed and associated true sources as a function of the true strain amplitude. The blue and red dots correspond to confirmed sources found from the analysis of the LDC1-4 data and the LISA-Taiji network, respectively. In each panel, an excess of red dots towards low values of $\mathcal{A}$ is visible, indicating that the LISA-Taiji network found more weak sources than LISA alone. }
    \label{fig:scatterplots}
\end{figure*}

The fact that there is no significant change in the PDFs in going from LISA to LISA-Taiji may appear surprising at first since one expects parameter estimation accuracy to improve due to the higher SNR of a source in the latter. However, one must also account for the fact that the network finds a larger number of weaker sources that, in general, have worse errors in the estimated parameters. This is supported by Fig.~\ref{fig:scatterplots}, which shows the scatterplot of the differences in the parameter values of confirmed and associated true sources as a function of the strain amplitude of the latter. As can be seen clearly, the scatterplot for network sources extends to lower amplitudes where the error in the parameters tends to be higher. 

\begin{figure*}
    \centering
    \includegraphics[width=\textwidth]{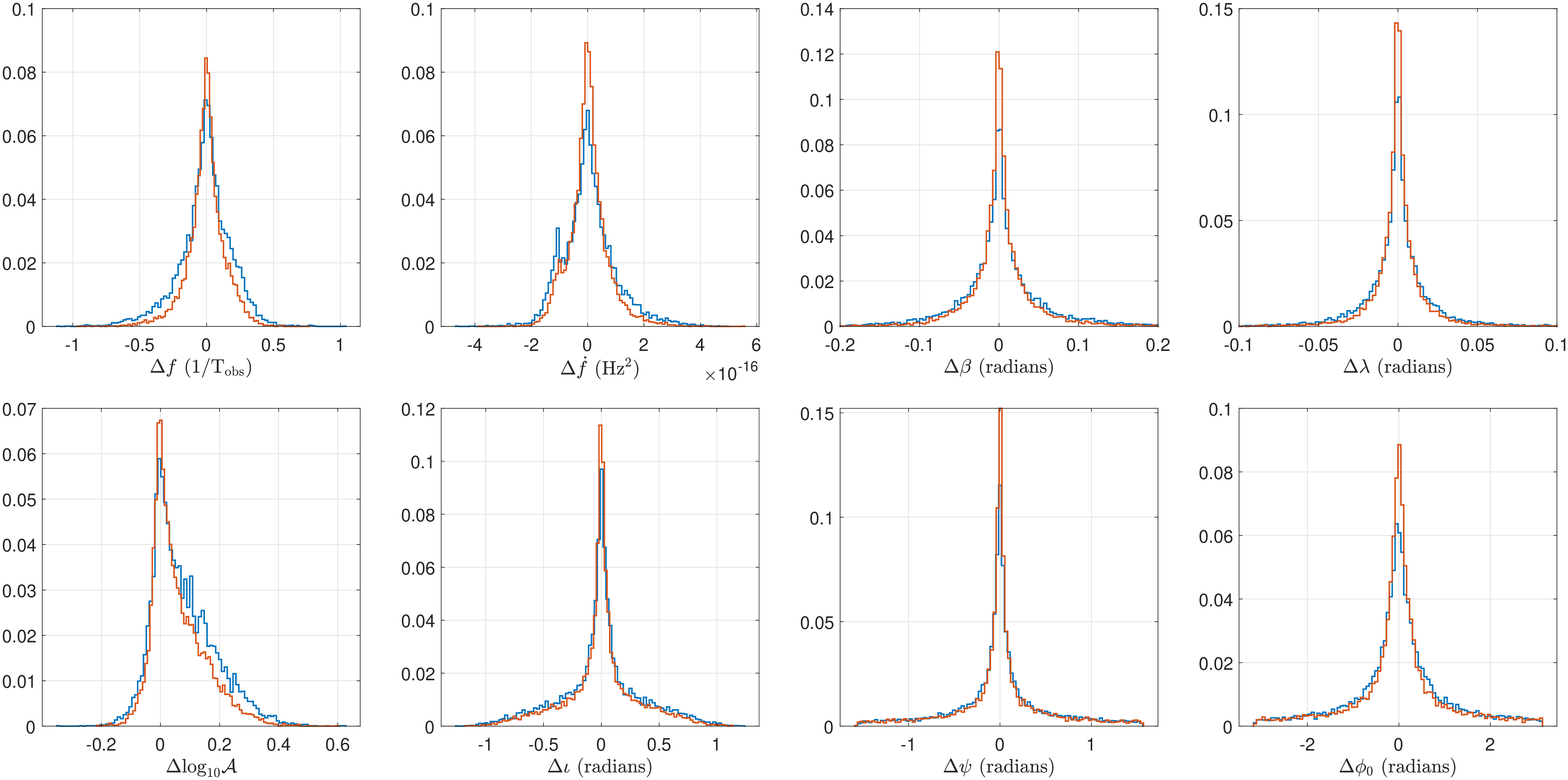}
    \caption{Estimated PDF of differences in the parameters of confirmed and associated true sources for only the subset of sources that were common to LISA (blue) and the LISA-Taiji network (red) analyses. A small fraction of outliers have been removed from the set of $\Delta\beta$ and $\Delta\lambda$ values for visual clarity: They constitute $1.7\%$($0.6\%$) and $1.1\%$($0.6\%$), respectively, for $\Delta\beta$ and $\Delta\lambda$ using single-detector (network) analysis. 
    The values of $\sqrt{2}\times\Sigma_{\rm{net}}/\Sigma_{\rm{sngl}}$ for the parameters are as follows. $f$: $1.0218$, $\dot{f}$: $1.0471$, $\beta$: $1.0388$, $\lambda$: $1.0697$, $\rm{log}_{10}\mathcal{A}$: $1.2003$, $\iota$: $1.2429$, $\psi$: $1.3649$, $\phi_0$: $1.3440$. }
    \label{fig:LISA+network-error}
\end{figure*}

For a fair comparison between the single-detector and network performance, we must look at the distribution of parameter  differences for only the confirmed sources that they have in common. Fig.~\ref{fig:LISA+network-error} compares the PDFs of the parameter differences for only this restricted set of sources and, indeed, it is seen clearly that they are all more concentrated towards zero difference in the case of the network. We have also listed in the caption of Fig.~\ref{fig:LISA+network-error}, the value of $\sqrt{2}\times \Sigma_{\rm net}/\Sigma_{\rm sngl}$ for each parameter where $\Sigma_{\rm net}$ and $\Sigma_{\rm sngl}$ are the standard deviations of the PDFs for the network and single-detector, respectively. Based on  Fisher information analysis of parameter estimation errors for a single source, in which the standard deviation of estimation error is inversely proportional to the SNR~\cite{Cutler:1994ys,Zhang:2020drf} and the factor of $\sqrt{2}$ accounts for the naive enhancement in SNR for a network of two detectors relative to a single one (all of them being identical with mutually independent noise), one expects this quantity to be close to unity. Indeed, we see that this is so for the subset of intrinsic parameters $f$, $\dot{f}$, $\beta$, and $\lambda$. However, it departs quite a bit from unity for all the extrinsic parameters. This result suggests that a Fisher information analysis confined to a single sources is not adequate for quantifying parameter estimation errors in the GB resolution problem.


\section{Conclusions}
\label{sec:conclusions}
We investigated the GB multisource resolution problem in the context of a network of comparable space-based GW detectors, namely, LISA and Taiji, using a previously developed iterative source subtraction pipeline, \fitalgoname, that was extended to analyze TDI combinations from multiple detectors. Our results provide a realistic assessment of the benefits expected from a network of space-based detectors that goes beyond Fisher information studies of parameter estimation errors for single GBs.

In the present work, we looked at detection and estimation performance of the LISA-Taiji network relative to LISA alone. There is a significant increase, by $\approx 75\%$, in the number of confirmed sources that can be extracted while maintaining approximately the same detection rate. This increase was obtained without changing any of the cuts (SNR and $R_{\rm ee}$) imposed on the identified sources or the cross-validation setup for primary and secondary runs. 

One of the key results we have reported is the comparison of the residual after subtraction of reported sources found in the network analysis with the ideal GB background of a single detector. We find that a network of space-based detectors will allow the GB background to be reduced substantially below the lowest level possible with a single detector. The PSD of the residual lies entirely below the ideal GB background obtained with a detection threshold of ${\rm SNR}_{\rm bg}^{\rm min}=7.0$. It coincides with the ${\rm SNR}_{\rm bg}^{\rm min}=5.0$ background over much of the frequency range in which the latter dominates over instrumental noise. Since the residual from subtracting GBs forms the noise background for non-GB searches, one expects that their performance will be significantly improved in the $\approx [0.5, 4]$~mHz band. Further studies need to be carried out to quantify this expectation.

Our study of parameter estimation errors brings forth two salient points that highlight the difference between a single-source and multisource problem. (i) There is no significant change in the distribution of the differences in parameter values of confirmed and true sources between a single-detector and a network of comparable ones. (ii) Fisher information based error analysis of single sources may not be adequate in the multi-source resolution problem since the expected scaling of errors with SNR is observed only for the subset of intrinsic parameters but not the extrinsic ones. While there is a clear explanation for (i), we do not understand the origin of (ii) yet. 

We have assumed simplified models for both the LISA and Taiji detectors, treating them as identical in design except for an angular separation in their heliocentric orbit. In future work, we will use more realistic models (the orbits and individual TDI combinations) for LISA and Taiji data to investigate the performance improvements for LISA-Taiji network. We will also take into account the uncertainty in prior knowledge of the PSDs of instrumental or GB background noise. This can be achieved, in principle, by a modest expansion of the search space for PSO by including the ratios of the noise PSDs in the same frequency band as free parameters to be optimized along with those of a single GB. We have further assumed that the LISA and Taiji data are collected over the same observation period. Since the GBs are persistent sources, and since the \fstat treats data from different detectors and TDI combinations additively, much of our analysis would go through, barring code modifications, even if the observation periods were non-overlapping. The inclusion of strongly evolving GBs could be an interesting exception here, if the observation periods are far apart, that will require a more careful approach.

\section*{Acknowledgements}
Zhang, Zhao and Liu are supported by the National Key Research and Development Program of China grants No. 2021YFC2203000 and No. 2020YFC2201400, the National Natural Science Foundation of China through Grant No. 12047501, the 111 Project under grant No. B20063, and Lanzhou City’s scientific research funding subsidy to Lanzhou University. Group activities are financially supported by the Morningside Center of Mathematics and a part of the MPG-CAS collaboration in low frequency gravitational wave physics. Partial support from the Xiandao B project in gravitational wave detection is also acknowledged. We are grateful to Prof. Shing-Tung Yau for his long term and unconditional support and Prof. Yun-Kau Lau for assembling our research group. We gratefully acknowledge the use of high performance computers at the Supercomputing Center of Lanzhou University and State Key Laboratory of Scientific and Engineering Computing, CAS.
 

\end{document}